\documentclass[12pt]{article}
\let\section=\subsection     \let\subsection=\subsubsection                
\newcommand{ \be }{\begin{equation}} 
\newcommand{ \ee }{\end{equation}} 
\newcommand{ \bea }{\begin{eqnarray}} 
\newcommand{ \eea }{\end{eqnarray}} 
\newcommand{ \la }{\langle} 
\newcommand{ \ra }{\rangle} 
\newcommand{ \eps }{{\varepsilon}} 
 
\usepackage{bm}
\usepackage{graphicx}

\begin{document}
\begin{center}
   {\large \bf ANISOTROPIC FLOW FROM AGS TO RHIC}\\[5mm]
   S.~A.~VOLOSHIN \\[5mm]
   {\small \it  Department of Physics and Astronomy, Wayne State University \\
   666 W. Hancock, Detroit, MI 48201, U.S.A. \\[8mm] }
\end{center}

\begin{abstract}\noindent
   The recent results on anisotropic flow in ultrarelativistic nuclear
   collisions  along with recent methodical developments and
   achievements in the understanding of the phenomena,
   are reviewed. The emphasis is given to the elliptic flow results. 
\end{abstract}

\section{Introduction}

In recent years, the subject of anisotropic flow in ultrarelativistic nuclear
collisions attracts an increasing attention of heavy ion community. 
One of the main reasons for that is the  sensitivity of
anisotropic flow, and in particular elliptic flow~\cite{Olli92}, 
to the evolution of the system at the very early times~\cite{Sorge97}.      

Anisotropic flow is defined as azimuthal asymmetry in
particle distribution with respect to the reaction plane (the plane
spanned by the beam direction and the impact parameter).
It is called {\it flow} for it is a collective phenomena, 
but it does not necessarily imply {\it hydrodynamic} flow.
It is convenient to characterize the magnitude of this asymmetry using
Fourier decomposition of the azimuthal distributions. Then the first
harmonic describes so-called {\it directed} flow, 
and the second harmonic corresponds to {\it elliptic} flow; 
non-zero higher harmonics can be also present in the distribution. 
The corresponding Fourier coefficients,
$v_n$ are used to quantify the effect~\cite{method}.

The two reasons for anisotropic flow are the original asymmetry in the
configuration space (non-central collisions !) and rescatterings.  
In a case of elliptic
flow the initial ``ellipticity'' of the overlap zone is usually
characterized  by the quantity 
$\eps=(\la y^2 - x^2 \ra) /(\la y^2 + x^2 \ra$, assuming the
reaction plane being $xz$-plane. 
With the system expansion the spatial anisotropy decreases. 
This is the reason for high sensitivity of elliptic flow to the
evolution of the system in the very early times~\cite{Sorge97,mpc},
of the order of 2--5~fm/c,
independent of the model. 
       
Due to the lack of space we will not discuss in detail all recent
developments regarding {\it directed} flow. 
Briefly mention a couple. 
In~\cite{wiggle} a very interesting {\it qualitative} prediction is
given: it is shown that the radial flow (isotropic expansion in 
the transverse plane) and an incomplete baryon stopping should lead to
a ``wiggle'' in the rapidity dependence of baryon directed flow;
$v_1(y)$ should change it's sign three times with rapidity!
This effect was also observed in a hydro model~\cite{thirdcomponent}.
Once found it would be a strong evidence for the space-momentum
correlations caused by radial flow. 
In a recent paper~\cite{MomConservation} the important question of the role
of the momentum conservation in directed flow measurements are
discussed. Concrete recommendations for the analysis have been worked out.

Recently, many new results regarding {\em elliptic} flow 
have been obtained in all
directions: experimental measurements, improving analysis methods, and
theoretical understanding of the underlying physics of the phenomena. 
I will try to mention the most important results in all three
directions, but concentrate mostly on the last two questions.
A more complete picture of the recent experimental results can be
found in~\cite{art}. 

\section{Improving the methods}  

A significant progress in theoretical description of anisotropic flow 
demands the accuracy in measurements. 
Thus the corresponding methods are
evolving in the directions of improving both, the
{\em statistical} uncertainties, and understanding {\em systematics} 
in the measurements.
For the first, improving on the statistical errors, we mention extensive
use of proper weights (the best would be $w \propto v_n(y,p_t)$),
which leads to the improvement of  the reaction plane resolution by
10\%-20\% and reduction of the statistical errors,
and using the scalar-product approach~\cite{scalarReport}. 
In the scalar-product method the flow is given by:
\begin{equation}       
v_n(y, p_t)     
= \frac{ \langle {\bf Q}_n \cdot {\bf u}(y, p_t) \rangle }       
  {2 \sqrt{ \langle {\bf Q}^a_n \cdot {\bf Q}^b_n \rangle } } \,,       
\end{equation}       
where ${\bf u} \equiv (\cos(n \phi),\sin(n \phi))$ 
is a unit vector associated with a particle of a given
rapidity and transverse momentum, and ${\bf Q}$ are flow vectors for
the ``full'' event and subevents ``a'' and ``b'':
\be
{\bf Q}_n =\sum_i {\bf u}_i,  
\ee
where sum is over all particles in an (sub)event. This method (which
is also easy to implement and analyze) in addition to the flow angle
takes into account the flow effects on the magnitude of the flow
vector and as a result gives smaller statistical errors. 
 
In is not possible to determine the reaction plane 
in the collision directly.
Therefore, any measurement of the anisotropy in particle production
with respect to the reaction plane is based on the measurements of
particle azimuthal correlations among themselves. Those correlations
to a different degree (depending on what exactly is analyzed) include
the contribution from the
correlations that are not related to the orientation of the
reaction plane (e.g. resonance decays), and often called
{\em non-flow} contribution. For a reliable interpretation of the
results the non-flow contribution should be estimated or, better, 
measured. 

\noindent
\hspace*{-0.8cm}
\includegraphics[width=8.4cm,angle=0]{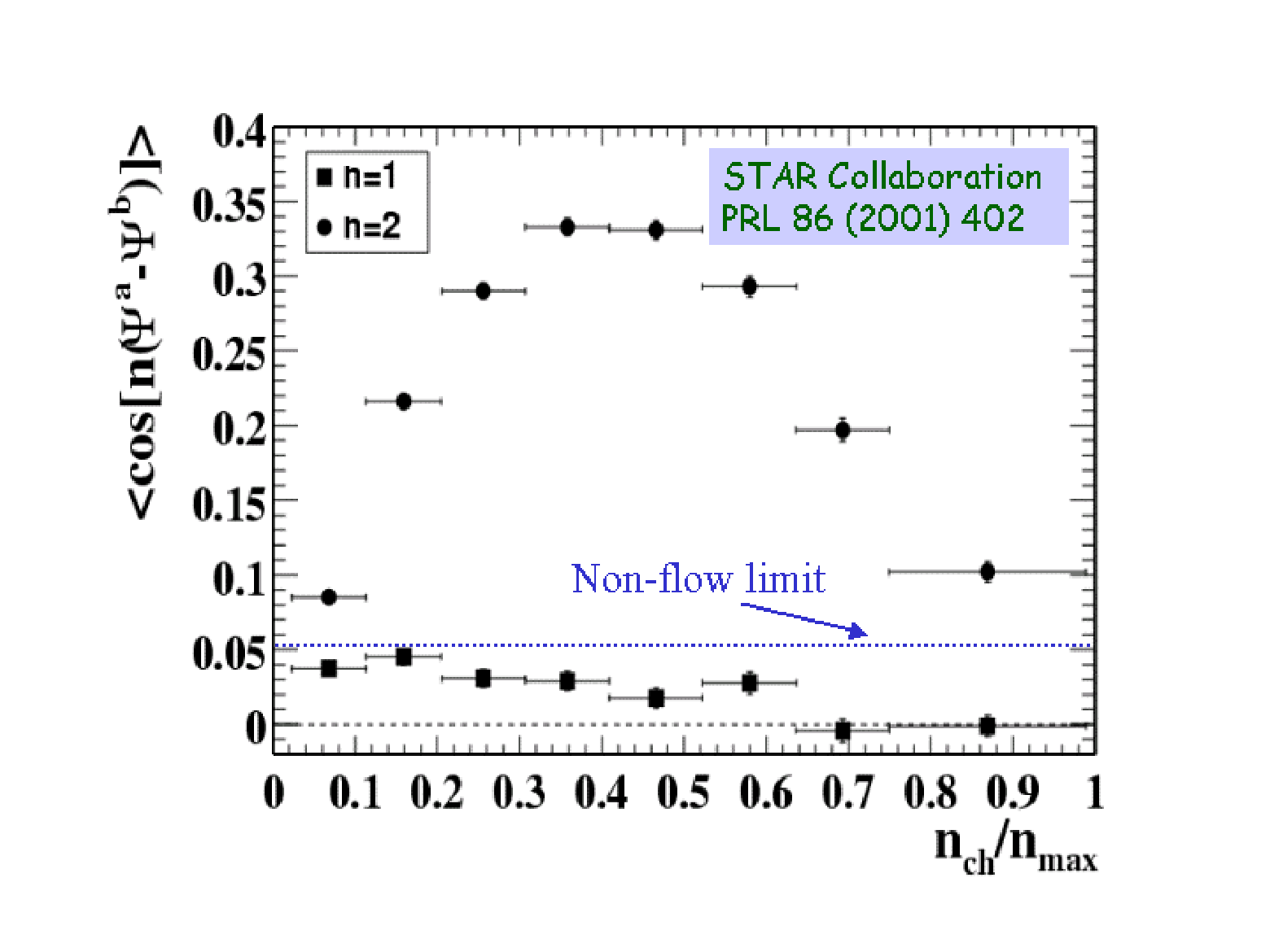}

\vspace*{-5.6cm}   
\hspace*{6.5cm}
\includegraphics[width=7.4cm,angle=0]{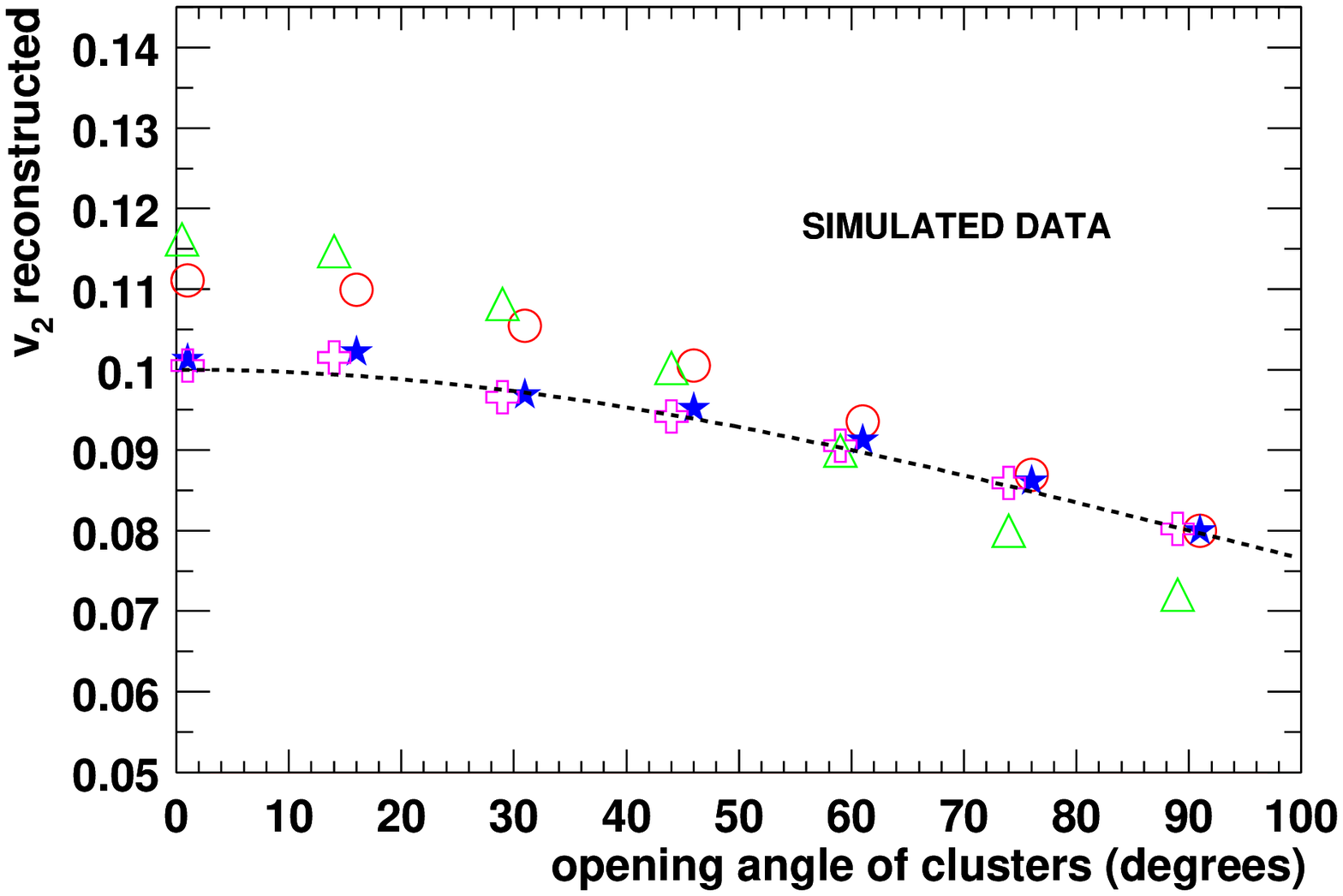}

\noindent
\vspace*{3mm}
\parbox{6.7cm}
{\footnotesize 
        Fig. 1: STAR~\cite{STARfirstflow}. 
Centrality dependence of the correlation
between subevent flow angles. 
}
\hspace*{2mm}
\parbox{6.7cm}
{\footnotesize 
   Fig.~2: Simulations~\cite{star_cumulants}. 
The results from 2- (triangles), 4- (stars) and 6-particle
(crosses) cumulants.
}

As an example, we discuss the estimate of non-flow contribution 
in STAR elliptic flow measurements~\cite{STARfirstflow}. 
An important observation for that is on the centrality dependence 
of the non-flow effects. 
The azimuthal correlation between two particles can be written as     
\be    
\la u_{n,1} u_{n,2}^* \ra \equiv \la e^{i n \phi_1} e^{- i n \phi_2} \ra    
=v_n^2 + \delta_n \,,    
\label{twonf}    
\ee    
where $n$ is the harmonic, and the average is taken over all pairs of     
particles.  
The $\delta_n$ represents the contribution to the pair correlation 
from non-flow effects.
Then, the correlation between two subevent flow    
angles is
\be   
\la \cos(2(\Psi_2^{(a)}-\Psi_2^{(b)})) \ra \propto M_{sub}(v_2^2 + g)
\propto     
M_{\rm sub} v_2^2 + \tilde{g},
\label{coscorr}   
\ee       
where $M_{sub}$ is the multiplicity for a sub-event.
Here, we have taken into account that the  strength of 
the non-flow correlations scale in inverse    
proportion to  the multiplicity: $g\propto 1/M_{\rm sub}$.   
What is important is that the non-flow contribution to $\la    
\cos(2(\Psi_2^{(a)}-\Psi_2^{(b)})) \ra $ is approximately independent    
of centrality.  The typical shape of 
$ \la   \cos(2(\Psi_2^{(a)}-\Psi_2^{(b)})) \ra $, see, for example,    
Fig.~1, is peaked at mid-central events due to    
the fact that for peripheral collisions, $M_{\rm sub}$ is small,     
and for central events, $v_2$ is small.  In the     
estimates~\cite{STARfirstflow} of the systematic errors, the authors set    
the quantity $\tilde{g}=0.05$.  
The justification for this value was    
the observation of similar correlations for the first and higher     
harmonics (it has been investigated up to the sixth harmonic).  One    
could expect the non-flow contribution to be of similar order of    
magnitude for all these harmonics, and model
simulations support this conclusion.  Given the value $\tilde{g} =     
0.05$, one simply estimates the contribution from non-flow    
effects to the measurement of $v_2$ from the plot of $\la    
\cos(2(\Psi_2^{(a)}-\Psi_2^{(b)})) \ra $ using Eq.(\ref{coscorr})
(see circle-point error-bars in Fig.6).    
The relative contribution of non-flow effects 
is largest for very central and very peripheral bins (where,
the reaction plane resolution is smallest!).

Anisotropic flow is a genuine multiparticle phenomena (which justifies    
use of the term {\em collective flow}).  
It means that if one considers many-particle correlations instead of    
two-particle correlations, the relative contribution of non-flow    
effects (due to few particle clusters) would decrease.  Considering    
many-particle correlations, one has to subtract the contribution from    
correlations in the lower-order multiplets and
use cumulants instead of simple correlation functions~\cite{olli_cumulants}.  
For example, correlating four particles, one gets    
\be    
\la u_{n,1} u_{n,2} u_{n,3}^* u_{n,4}^* \ra    
= v_n^4 + 2 \cdot 2 \cdot v_n^2 \delta_n + 2 \delta_n^2 \,.    
\label{fournf}    
\ee    
In this expression, two factors of ``2'' in front of the middle term    
correspond to the two ways of pairing (1,3)(2,4) and (1,4)(2,3) and     
account for the possibility to have non-flow effects in the first    
pair and flow correlations in the second pair and vice versa.  The    
factor ``2'' in front of the last term is due to the two ways of    
pairing.  The pure four-particle non-flow correlation is omitted     
from this expression (see below on the     
possible magnitude of such a contribution).  
If one subtracts from the expression (\ref{fournf}) twice the    
square of the expression (\ref{twonf}), one is left with only the     
flow contributions     
\be    
\la \la  u_{n,1} u_{n,2} u_{n,3}^* u_{n,4}^* \ra \ra        
\equiv    
\la u_{n,1} u_{n,2} u_{n,3}^* u_{n,4}^* \ra    
-2 \la u_{n,1} u_{n,2}^* \ra ^2    
= -v_n^4 \,,    
\ee
where the notation $\la \la ... \ra \ra$ is used for 
the {\em  cumulant}.  
A very elegant way of calculating cumulants in flow analysis with the
help of the generating function is proposed in~\cite{olli_cumulants}.
The simulations~\cite{star_cumulants}, see Fig.~2, confirm that 
using 4-particle
cumulants reliably removes non-flow contributions. It also shows that
even in the presence of genuine 4-particle correlations (due to
clusters decaying into 4-particles, such were introduced into
simulations) 
those correlations are combinatorially suppressed compared to
real flow correlations and there is no real need for use of higher order
cumulants in flow analysis.  
   
The high precision results available 
in modern high statistics and large
acceptance experiments  become sensitive   
to another effect usually neglected in flow analysis, namely,    
event-by-event {\em flow fluctuations}.  The latter can have two   
different origins: ``real'' flow fluctuations -- fluctuations    
at fixed impact parameter and fixed multiplicity (see, for example,   
~\cite{kodama,raju}), and impact parameter variations among events from    
the same centrality bin in a case where flow does not fluctuate at fixed    
impact parameter.     
Note that these fluctuations affects
any kind of analysis, including the ``standard'' one based on   
pair correlations.  The reason is that any flow measurements are    
based on correlations between particles, which     
are sensitive only to certain moments of the distribution in $v_2$.   
In the pair correlation approach with the reaction plane determined    
from the second harmonic, the correlations are proportional   
to $v^2$.  Averaging over events gives $\la v^2 \ra$,   
which in general is not equal to $\la v \ra^2$.     
The 4-particle cumulant method involves the   
difference between 4-particle correlations and (twice) the square of the   
2-particle correlations.  It is usually assumed that this   
difference comes from non-flow correlations.    
Note, however, that this    
difference ($\la v^4 \ra - \la v^2 \ra^2 \ne 0$) could be   
due to flow fluctuations.  Let us consider an example where the   
distribution in $v$ is flat from $v=0$ to $v=v_{\rm max}$.   
Then, a simple calculation would lead to the ratio of the flow values   
from the standard 2-particle correlation method and 4-particle cumulants   
as large as $\la v^2 \ra^{1/2}/(2\la v^2 \ra^2 - \la v^4 \ra)^{1/4}   
= 5^{1/4}\approx 1.5$.

\section{The physics of elliptic flow}

Many important developments in this area: 
better understanding of the transverse momentum dependence
of anisotropic flow in low $p_t$ region with the help of 
the ``blast wave'' model~\cite{blast,STAR_pid}, 
attempts~\cite{GVW01} to describe $v_2(p_t)$ in high $p_t$ region accounting 
for the parton energy loss -- ``jet quenching'',
calculation in a parton cascade model~\cite{molnar}, 
analysis of the anisotropies in Color Glass Condensate~\cite{raju}, 
and detailed analysis of the elliptic flow in  
the hydro models~\cite{kolb,Shuryak99,hirano}.

A very interesting development in this field is an attempt to calculate
elliptic flow in Color Glass Condensate -- classical field approach to
describe ultrarelativistic nuclear collisions.
One of the important consequences in this approach~\cite{raju} 
is strong event-by-event fluctuations in $v_2$. 
As it has been already discussed, such fluctuations would manifest 
themselves in the difference of the
flow results derived from 2- and 4-particle correlations.

\subsection{Transverse momentum dependence}

I would like to mention here what is usually referred to as ``hydro
inspired'', ``blast wave'' or ``expanding shell'', models.
Such models consider particle production from a thermal source in a
form of an expanding shell with the radial expansion velocity having
some azimuthal modulation. The case of directed flow was discussed
in~\cite{shell_directed}. That model was used to fit E877 data and
gave quite reasonable results~\cite{shell_directed}. The model was
generalized for the elliptic flow case in~\cite{blast}. Later
it was used to fit STAR data and was further
generalized for the case of the elliptic shape
shell~\cite{STAR_pid}. In this approach:
\begin{equation} 
  v_2(p_t)= 
  \frac{  \int_0^{2\pi} d\phi_b 
    \cos(2 \phi_b) 
    I_2(\alpha_t) K_1 (\beta_t) 
    (1+2 s_2 \cos(2\phi_b))} 
  {  \int_0^{2\pi} d\phi_b
    I_0(\alpha_t) K_1 (\beta_t) 
    (1+2 s_2 \cos(2\phi_b))}, 
  \label{modblastwave}
\end{equation} 
where $I_0$, $I_2$, and $K_1$ are modified Bessel functions, and where
$\alpha_t(\phi_b)=(p_t/T_f)\sinh(\rho(\phi_b))$, and 
$\beta_t(\phi_b)=(m_t/T_f)\cosh(\rho(\phi_b))$. 
The assumptions of this model are 
boost-invariant longitudinal expansion and freeze-out at constant 
temperature $T_f$ on a thin shell, which expands with a 
transverse rapidity exhibiting a second harmonic azimuthal modulation, 
 $\rho(\phi_b)=\rho_0 +\rho_a \cos(2 \phi_b)$. 
Here, $\phi_b$ is the azimuthal angle (measured with 
respect to the reaction plane) of the boost of the source element 
on the freeze-out hyper-surface~\cite{blast}, and $\rho_0$ and
$\rho_a$ are the 
mean transverse expansion rapidity ($v_0 = \tanh(\rho_0)$) and the 
amplitude of its azimuthal variation, respectively. 
In Fig.~3, the fit to the minimum-bias data with $s_2 = 0$ 
is shown as the dotted lines. 
The relatively poor fit led the authors to introduce a spatially 
anisotropic freeze-out hyper-surface, 
with one extra 
parameter, $s_2$, describing the variation in the azimuthal density of 
the source elements, $\propto 2 s_2 \cos(2 \phi_b)$. 
This additional parameter leads to a good description of the data, 
shown as the solid lines in Fig.~1. 
A positive value of the $s_2$ parameter would mean that there are more
source elements moving in the direction of the reaction plane.
The model predicts a specific dependence of the elliptic flow on the
particle mass.
This mass-dependent effect is larger for lower temperatures 
($T_f$) and larger transverse rapidities ($\rho_0$).

The behavior of $v_2(p_t)$ at {\em large transverse momenta} is also very
interesting. One of the possibilities is that the anisotropy at such 
transverse momenta is due to path length
dependent nuclear modification of the parton fragmentation function
(jet quenching). High $p_t$ parton produced in the direction of long
axis of the overlapping region exhibits more inelastic (in addition to
elastic) collisions than that emitted along the short axis. It results
in smaller probability to fragment into high $p_t$ hadron. 
The effect depends on the density of the media and thus the observed
anisotropy could serve as a measure on this very density~\cite{GVW01} 
(and features of the energy loss itself). The transverse momenta, where
$v_2(p_t)$ saturates could also help in
understanding the origin of the particles in the region of 2--5~GeV/c:
do they acquire their transverse momentum due to multiple scattering
or they come from a fragmentation of even higher $p_t$ parton?   
The preliminary STAR data~\cite{kirill}, Fig.~4, support the idea of flow
saturation at high $p_t$. 

\hspace*{-0.9cm}   
\includegraphics[width=7.4cm,angle=0]{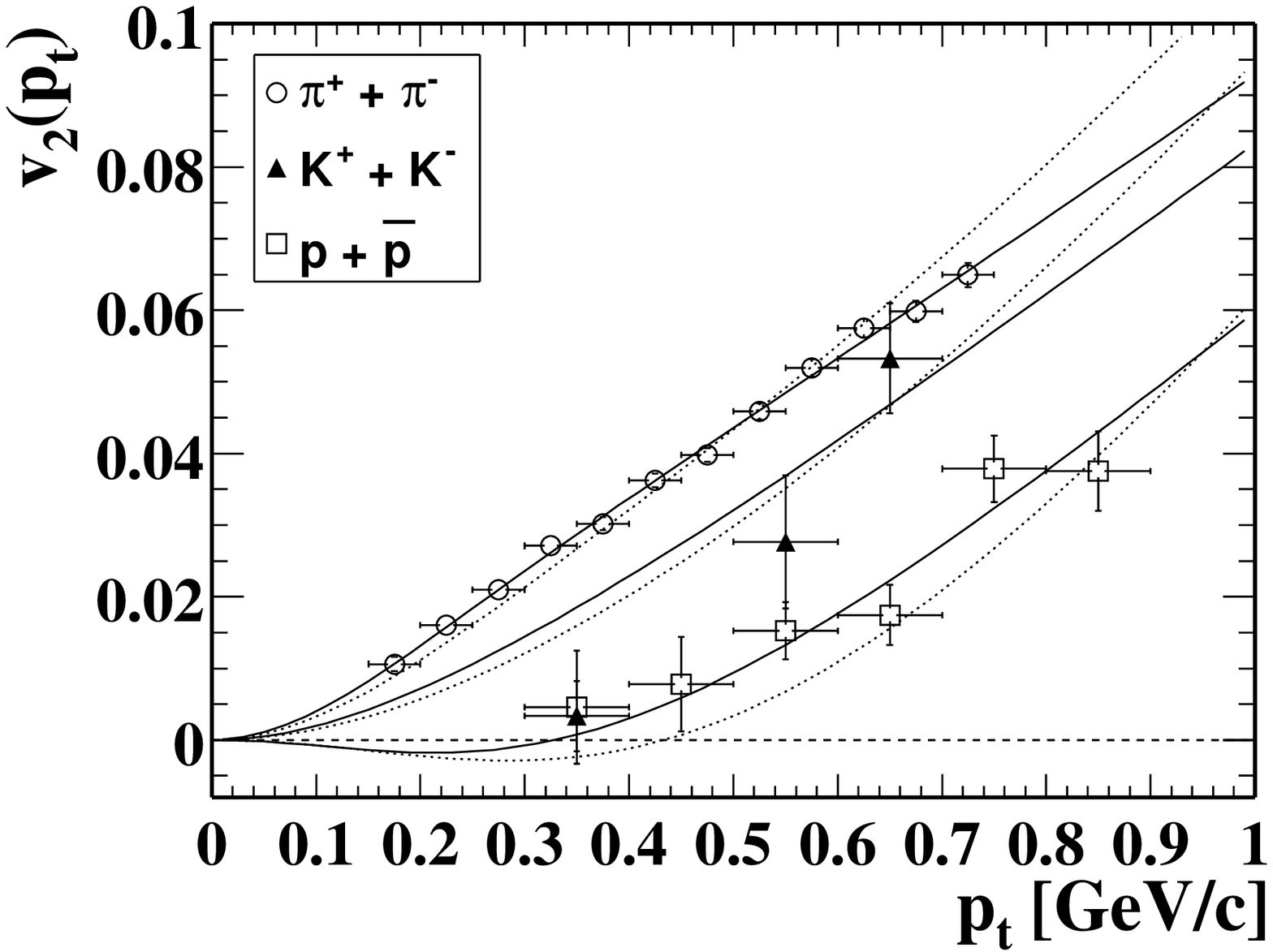}
\includegraphics[width=7.4cm,angle=0]{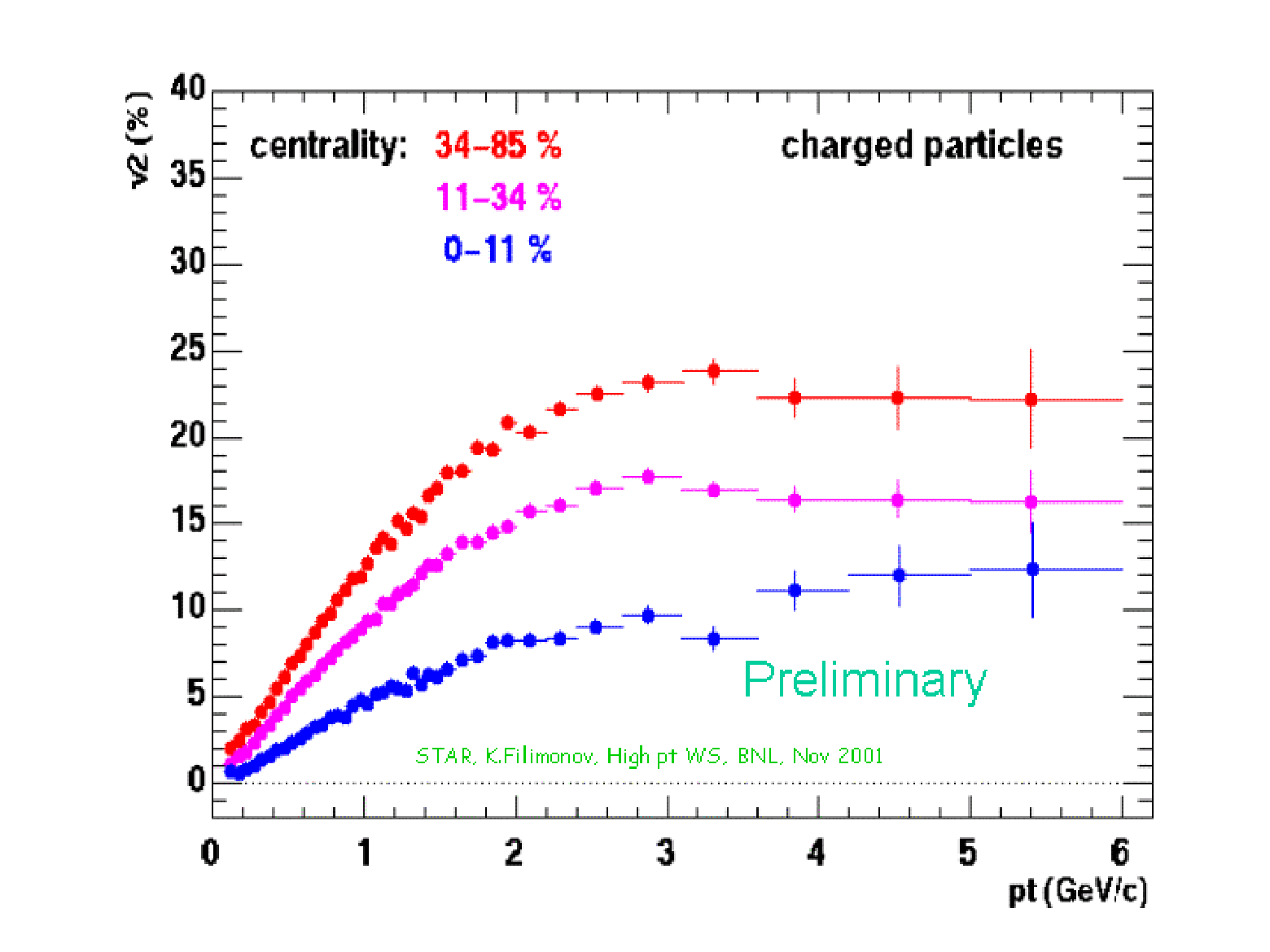}\\
\parbox{6.5cm}
{\footnotesize 
        Fig.~3: STAR~\cite{STAR_pid}. $v_2(p_t)$ with fits 
   (Eq.~\ref{modblastwave}) including $s_2$ parameter (solid lines) and
   without (dashed lines).
}
\hspace{0.9cm}   
\parbox{6.5cm}
{\footnotesize 
        Fig.~4: STAR~\cite{kirill}. $v_2(p_t)$ for charged particles. 
}

\subsection{Hydro and low density limits}

The values of elliptic flow measured at RHIC are  comparable to that
in hydrodynamic models. There was a clear disagreement at lower
energies. 
As it has been  mentioned in the introduction,
for elliptic flow one needs rescatterings. Denser the matter and more
rescattering means higher elliptic flow. The current understanding is
that in limit of zero mean free path, the {\em hydro limit}, one gets the
largest possible values of elliptic flow.

Interesting that the flow values obtained in parton
cascade calculations~\cite{molnar} at different transport opacities could in
principle significantly exceed the flow values from hydro
calculations. It raises a question about validity of the assumption
that the largest values of flow can be reached in hydro model.
Two lines shown in Fig,~6 correspond to the results of 
this model for two values of transport opacity, corresponding to
35 (upper line) and 13 times higher gluon density as given by
HIJING model. Note, however, that this opacity was calculating assuming
1~mb gluon transport cross section and assuming the hadronization
picture when the number of gluons equals the number of hadrons.
If one would, for example, consider a system as a constituent quark
gas, one would have to increase the cross section approximately by 3
times and quark density by approximately 2 to 3 times. That would give
the opacities very close to that needed to describe the data.  

\hspace*{-0.9cm}   
 \includegraphics[width=7.4cm,angle=0]{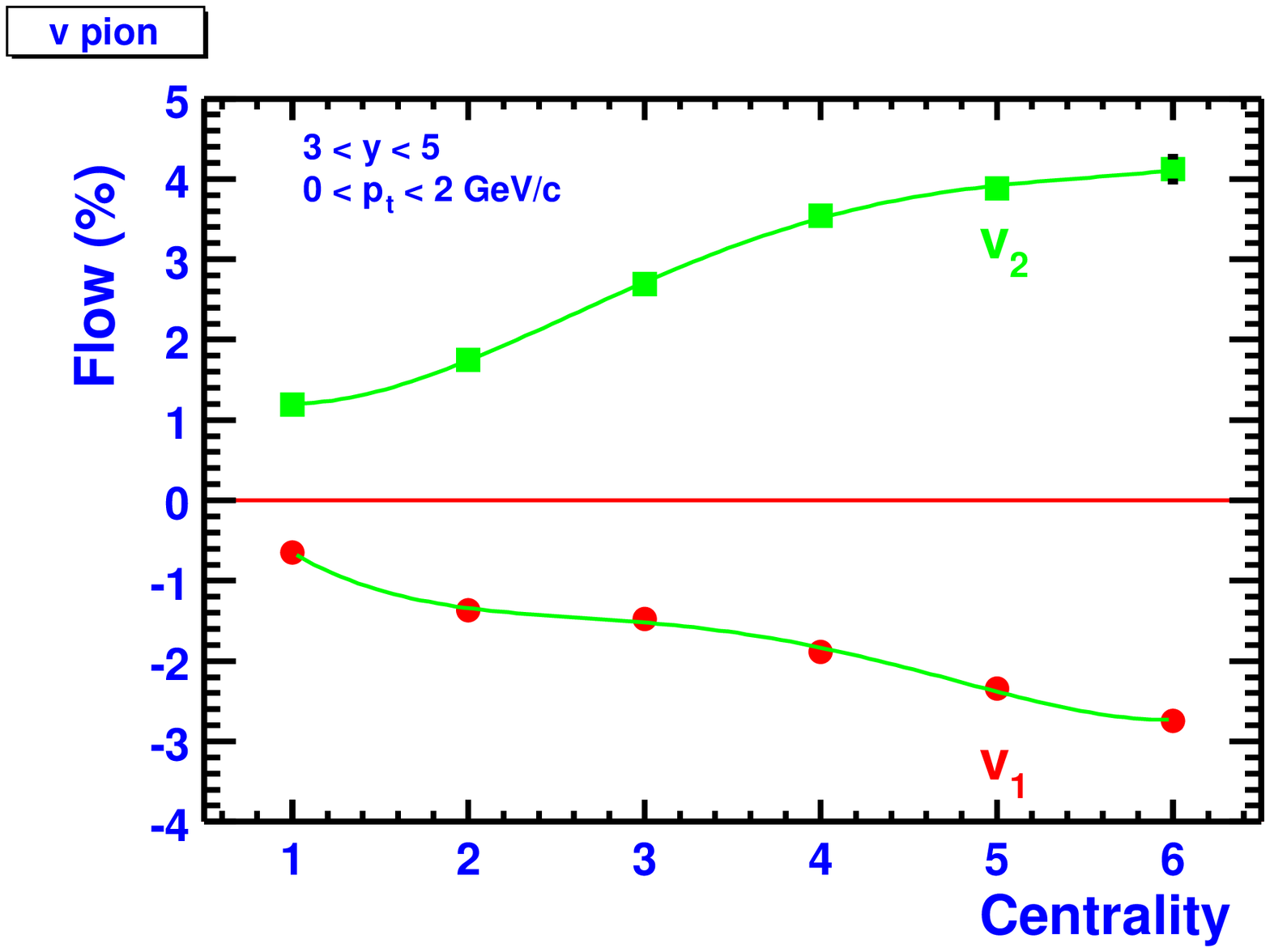}
\includegraphics[width=7.4cm,angle=0]{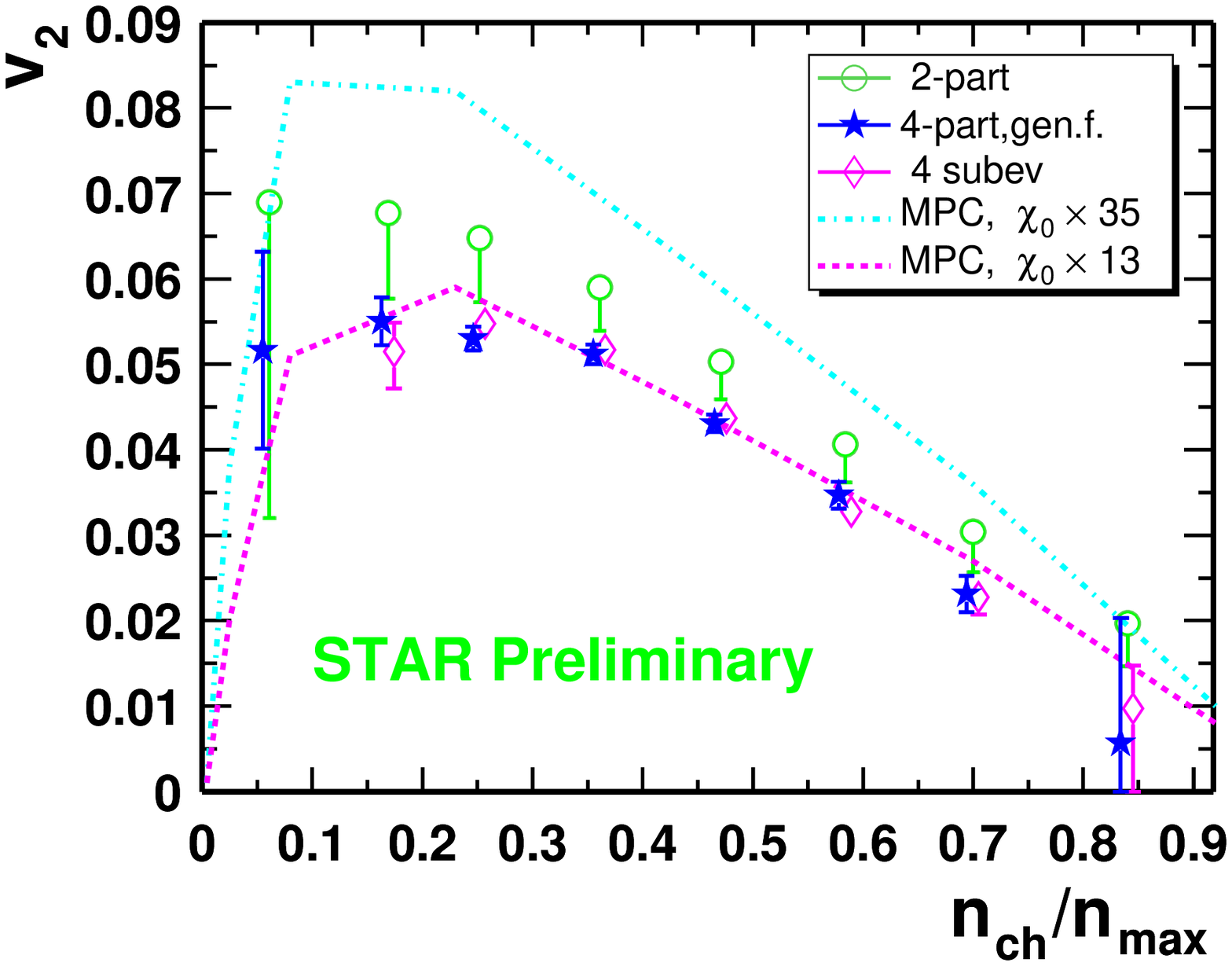}\\
\parbox{6.5cm}
{\footnotesize 
   Fig.~5: Preliminary NA49 results~\cite{na49data} on flow 
centrality dependence. 
}
\hspace{0.9cm}   
\parbox{6.5cm}
{\footnotesize 
        Fig.~6: STAR~\cite{star_cumulants}. $v_2$ centrality
 dependence from 2- and 4-particle correlations. Lines are the
 predictions from a parton cascade model\cite{molnar} (see text).
}

In the hydro limit elliptic flow
is  basically proportional to the original spatial ellipticity 
of the
nuclear overlapping region~\cite{Olli92,kolb,Shuryak99}, $v_2\propto \eps$.
In the opposite limit, 
usually called the {\em low density limit}~\cite{Heiselberg,vpPLB},
elliptic flow depends also on the particle density 
in the transverse plane: $v_2\propto \eps \;dN/dy \;/S$, where $S$ is the
area of the overlapping zone. It results in a different
centrality dependencies of the elliptic flow in these two limits.
The comparison of the results on elliptic flow from this point of view
was first done in~\cite{vpPLB}. In this picture, the transition 
to the deconfinement
would lead to some wiggles in $v_2/\eps$ dependence,
(``kinks'')~\cite{Sorge99,Heiselberg,vpPLB}).

\hspace*{-0.9cm}   
   \includegraphics[width=7.2cm,angle=0]{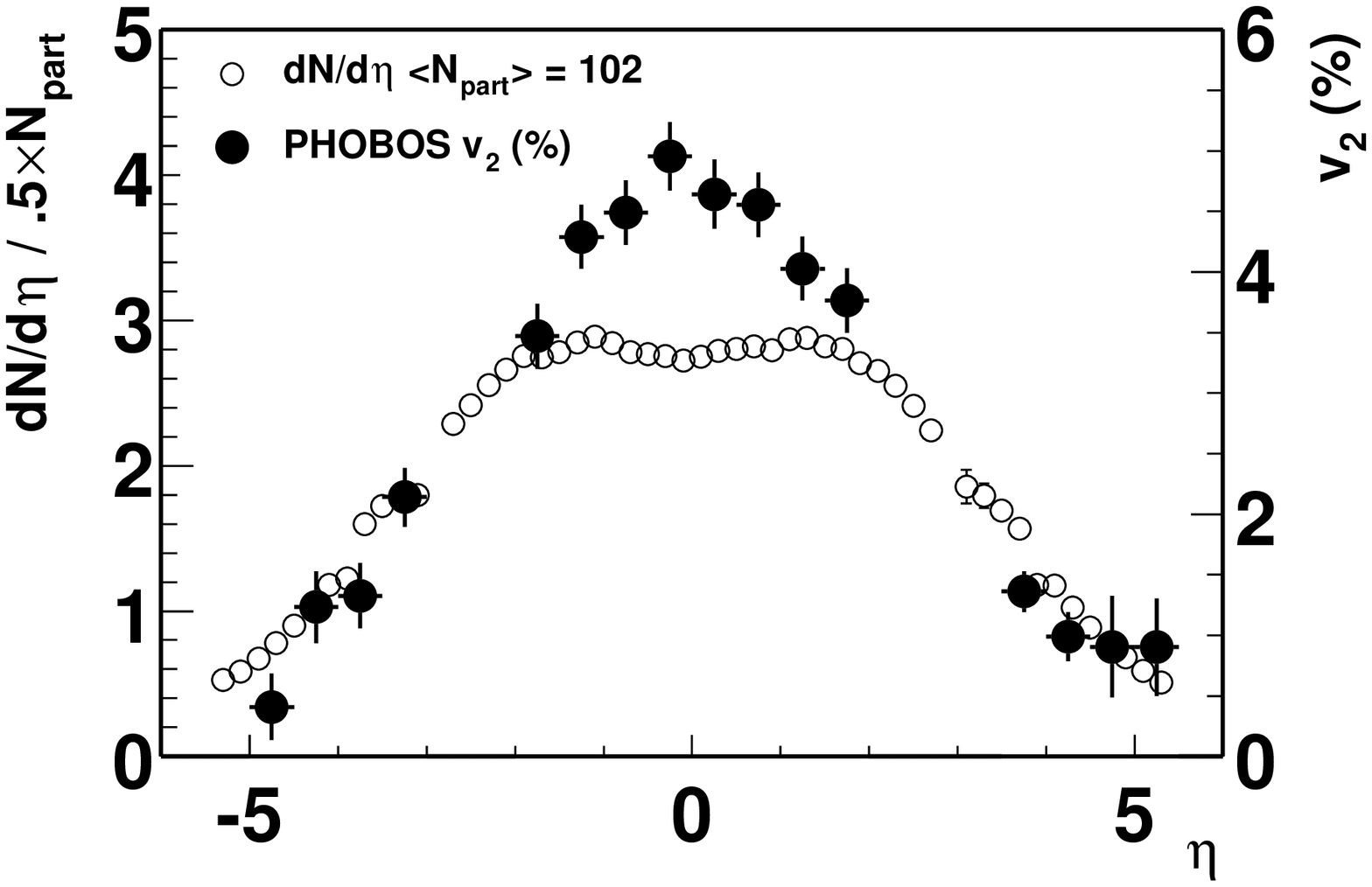}
   \includegraphics[width=7.6cm,angle=0]{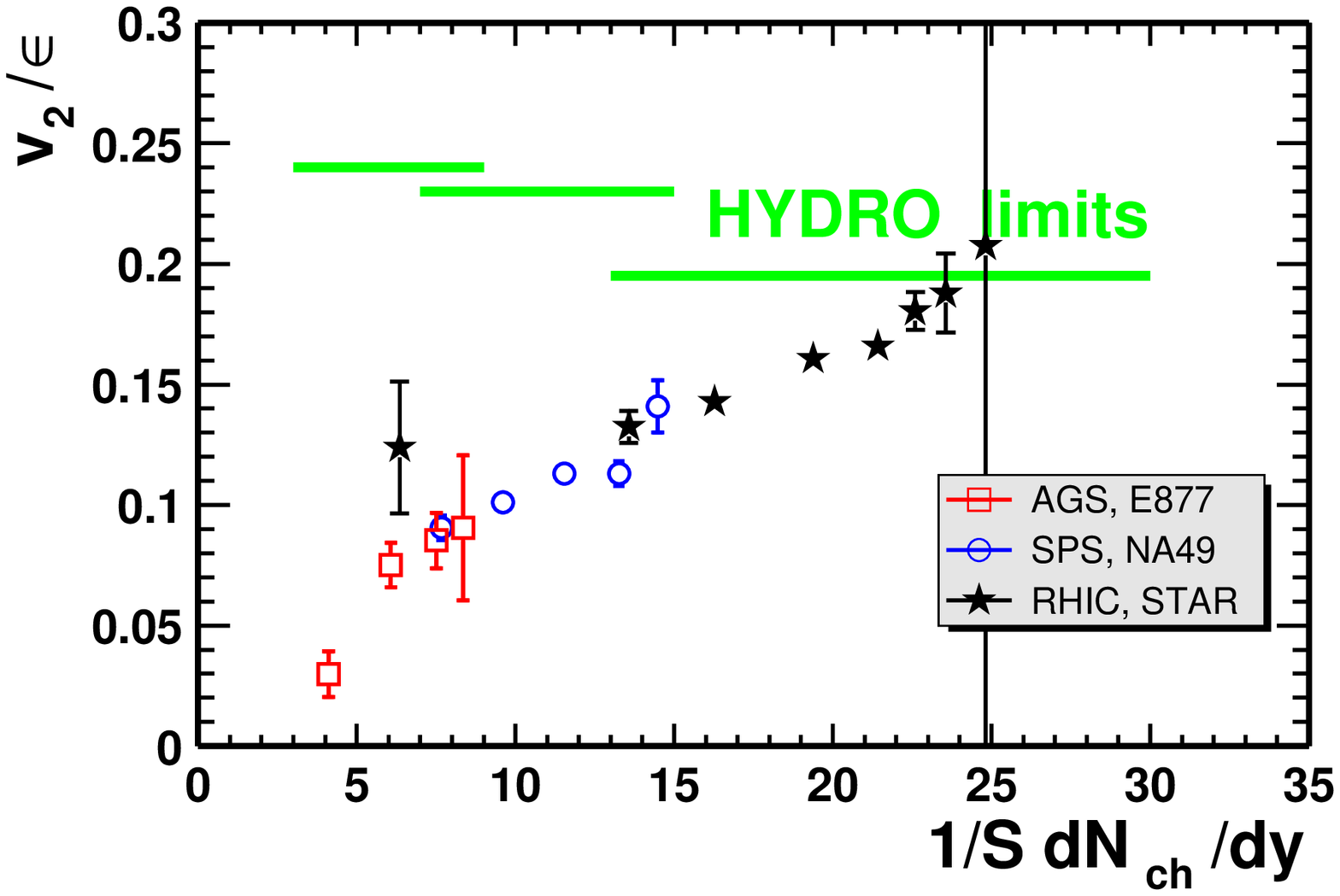}\\
\parbox{6.5cm}
{\footnotesize 
   Fig.~7: PHOBOS results on $v_2(\eta)$ overlayed with
   $dN/d\eta$~\cite{steinberg}.
}
\hspace{0.9cm}   
\parbox{6.5cm}
{\footnotesize 
        Fig.~8: $v_2/\eps$ as a  function of particle rapidity density
   for three colliding energies. Hydro limits taken from~\cite{kolb}.
}

One indication that at RHIC energies flow is still proportional to the
particle density can be seen from Fig.~7 (taken from~\cite{steinberg})
which show that $v_2(\eta)$ closely follows $dN_{ch}/d\eta$.
The 3D hydro calculations~\cite{hirano} also cannot describe the
pseudorapidity dependence of elliptic flow, once more
indicating that the hydro
description could be not correct in spite of the large values of $v_2$ 
measured at RHIC.   

Taking the recent (year 1, $\sqrt{s_{NN}}=130$~GeV) 
STAR results on elliptic flow 
from 4-particle cumulants~\cite{star_cumulants} (no non-flow
effects) and plotting them  along preliminary NA49~\cite{art} and E877 results,
Fig.~8, also suggests that even at RHIC energies elliptic flow
continue to rise with particle density. (Note possible systematic
errors in Fig.~8 of the order of 10-20\% due to uncertainties 
in centrality measurements. However, this uncertainty does not alter the
general trend).   
Decisive measurements could be
already at full RHIC energies of $\sqrt{s_{NN}}=200$~GeV. If $v_2$
would continue to rise with particle density (which increases for
about 15\% in this energy range), it could give difficulties
to hydrodynamic interpretation.

{\bf Acknowledgments}. The author thanks the organizers of the very
productive and exciting workshop
for the invitation. Numerous discussions with Art Poskanzer are
greatly appreciated.


\end{document}